\documentclass{article}
\usepackage{authblk}
\usepackage[utf8]{inputenc}
\usepackage{amsmath,amssymb}
\DeclareMathOperator\arctanh{arctanh}
\usepackage{graphicx}
\usepackage[
backend=biber,
style=alphabetic,
sorting=ynt
]{biblatex}

\title{Compression of a pressurized spherical shell by a spherical or flat probe}
\author[1]{\'Etienne Couturier} \author[2]{Dominic Vella} 
\author[3]{Arezki Boudaoud}

\affil[1]{Laboratoire Mati\`ere et Syst\`emes Complexes, Universit\'e de Paris CNRS UMR 7057, 10 Rue Alice Domont et L\'eonie Ducquet, 75205 Paris Cedex 13, France}
\affil[2]{Mathematical Institute, University of Oxford, Woodstock Rd, Oxford, OX2 6GG, UK}
\affil[3]{LadHyX, CNRS, Ecole polytechnique, IP Paris, 91128 Palaiseau Cedex, France}
\begin{document}

\maketitle
\abstract{Measuring the mechanical properties of cells and tissues often involves indentation with a sphere, or compression between two plates. Different theoretical approaches have been developed to retrieve material parameters (e.g.~elastic modulus) or state variables (e.g.~pressure) from such experiments. Here, we extend previous theoretical work on indentation of a spherical pressurized shell by a point force to cover indentation by a spherical probe or a plate. We provide formulae that enable the   modulus or pressure to be deduced from experimental results with realistic contact geometries, giving different results that are applicable depending on pressure level. We expect our results to be broadly useful when investigating biomechanics or mechanobiology of cells and tissues.
} %end of abstract
\maketitle
\section{Introduction}
In 1932, K.~Cole \cite{cole1932surface} introduced the mechanical compression of a sea urchin egg cell between two plates as a way to probe cell mechanical properties. This technique is now widely applied to both living and non-living  objects ranging from microcapsules \cite{bando2013deformation} and cell nuclei \cite{Goswami.2020} to single animal \cite{hiramoto1963mechanical,yoneda1964tension}, plant \cite{blewett2000micromanipulation} or, yeast \cite{zhang1999modelling} cells, as well as animal embryos \cite{davidson1999measurements} and multicellular spheroids \cite{Stirbat.2013} to name a few examples.

As might be expected with such a range of applications, nuances and differences of protocol have arisen, particularly with regard to the type of loading that is used. Different types of loading have different advantages and disadvantages: parallel plate compression is generally used to  probe cells globally, while atomic force microscopy (AFM) has been used to make more local measurements \cite{Alcaraz.2017,Milani.2013}. In the first implementations of AFM, the cantilevers had relatively sharp pyramidal tips, which could damage cells. Spherical beads used as tips appear to be less invasive and to generate less damage than sharp tips; they also make it possible to modify the range of applied stress by changing the bead diameter\cite{mahaffy2000scanning}.

Despite differences in loading protocol, all experimental approaches are similar in that they yield data for the force as a function of plate/tip displacement (or vice versa). Interpreting such force--displacement curves is therefore key, yet, since these curves depend on sample geometry and structure, the deduction of material parameters involves the use of mechanical models. In several cases, the sample can be approximated as a spherical thin shell, which might correspond to the cortex of an animal cell, the cell wall of a plant cell, or a peripheral stiff cell layer. Moreover, this shell is often pressurized from inside, whether that pressure arises from cytoplasmic pressure in the case of a single cell or pressure from inner cells in a multicellular spheroid. Here we aim to establish a suitable model for the compression of a pressurized spherical shell by a rigid flat or spherical probe. The first related work in this regard concerned compression by a flat plate, but  neglected the  effect of bending stiffness within the shell \cite{feng1973contact}. This model was later corrected for the incompressibility of the cell fluid content \cite{lardner1980compression}, though the effect of bending stiffness was still neglected. 

Moduli estimated by compression between two plates may significantly differ from moduli estimated with an AFM tip due to the difference of scale at which cells are probed (\cite{touhami2003nanoscale},\cite{stenson2011determining},\cite{thomas2010measuring}). A model for a point-like probe indenting a pressurized shell was developed in \cite{vella2012indentation,Vella.2012xk6}, yielding analytical formulae to retrieve both the inner pressure and Young's modulus of the shell. Nevertheless, a model applicable to  probe sizes between a point (the idealized AFM tip) and a flat plate is still missing. Here we establish such a model: we use an efficient formulation for the nonlinear contact problem for a shallow shell without friction that was introduced in \cite{audoly2010elasticity} and provide an analytical solution of the linearized problem that is valid for small displacement. Beyond this  we solve the nonlinear boundary value problem  numerically.

\begin{figure}
\includegraphics[width=0.6\linewidth]{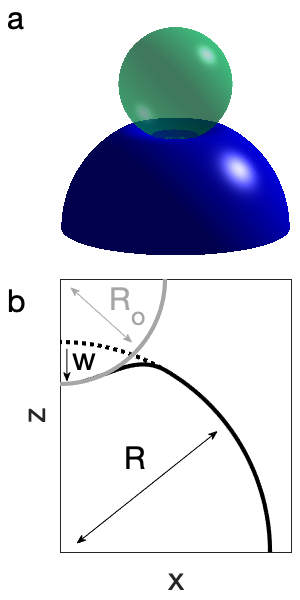}
\label{Figure_1}
\caption{Geometry of the problem: a pressurized spherical shell  in contact with a spherical probe. a.  Three-dimensional schematic drawing showing the indenter (green, partially transparent sphere) indenting the shell (blue, deformed sphere). b. Cross-section and notations showing the pressurized sphere before indentation (dotted black curve), the pressurized sphere after indentation (continuous black curve) and the rigid spherical probe (grey curve). Here, $w$ is the normal displacement of the probe, while $R$  is the radius of the pressurized sphere and $R_o$ the radius of the rigid sphere (the indenter).}
\end{figure}

\section{Formulation and numerical solutions}
\begin{figure}
\includegraphics[width=0.9\linewidth]{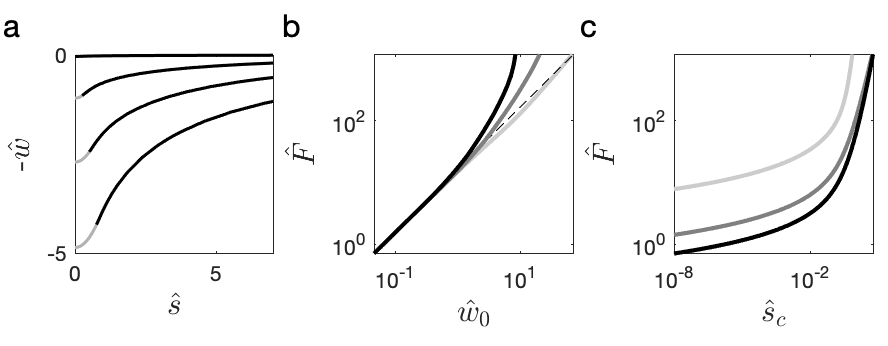}
\label{Figure_2}
\caption{Numerical solutions for the contact of a pressurized spherical shell with a rigid spherical or flat probe. a. Dimensionless displacement $\hat{w}$ as a function of the dimensionless coordinate $\hat{s}$ for $\hat{p}=1$ and $R/R_o=1$. The grey (resp. black) part of the line correspond to the contact zone (resp. free zone).
b. Dimensionless force as a function of the dimensionless displacement at the origin $\hat{w}_0$. The dashed line is a linear fit for small compressions $\hat{w}\lesssim1$. c. Dimensionless force as a function of the dimensionless coordinate,  $\hat{s}_c$, at which contact between the shell and the probe is lost. In (b) and (c), the dimensionless pressure  $\hat{p}=8$; black, dark gray, and gray lines show results for  $R/R_o=0$ (a plane), $R/R_o=1$, and $R/R_o=10$, respectively.}
\end{figure}

We consider a rigid spherical probe of radius $R_o$ (the planar case is recovered by taking $R_o=\infty$) pressed against a spherical shell of radius $R$, thickness $h$, Poisson ratio $\nu$, and Young's modulus $E$, that is inflated by a pressure $P$. We assume axisymmetry and express all variables as functions of the curvilinear coordinate $s$ (along the meridional direction). The periphery of the contact region between the probe and the shell is defined by its curvilinear coordinate $s_c$. Following \cite{audoly2010elasticity}, we formulate the equations using the rotation $\Psi=-\mathrm{d} w/\mathrm{d} s$, 
$w$ being the displacement normal to the sphere,  the meridional (the $ss$ component of) membrane stress  $\Sigma$, and the total reaction force  integrated up to a disk of radius $s$, denoted $\phi(s)$ (see Figure \ref{Figure_1}b). The total contact force is $F=\phi(s_c)$. The governing equations are given by the compatibility equation \cite{audoly2010elasticity}:
\begin{equation}
\frac{3s\Sigma^\prime+s^2\Sigma^{\prime\prime}}{E h}+s\frac{\Psi}{R}+\frac{\Psi^2}{2}=0,\label{eq_Sigmadim}
 \end{equation}
 and by the vertical force balance
 \begin{equation}
s\Sigma\left(\frac{s}{R}+\Psi\right)-\frac{Eh^3}{12(1-\nu^2)}\left(\frac{d(s\Psi^\prime)}{ds}+\frac{\Psi}{s}\right)=p\frac{s^2}{2}-\frac{\phi}{2\pi}.\label{eq_Psidim}
 \end{equation} 
These equations are complemented by the boundary conditions at the edge of the contact region, $s=s_c$, which require continuity of rotation, curvature and stress at the edge of the contact set, i.e.
\begin{align}
  \Psi(s_c)&=-s_c\left(\frac{1}{R}+\frac{1}{R_o}\right),\ \Psi^\prime(s_c)=-\left(\frac{1}{R}+\frac{1}{R_o}\right),\nonumber \\ 
  \Sigma^\prime(s_c)&=\frac{Eh\sqrt{\epsilon}s_c}{8R}\left(1-\left(\frac{R}{R_o}\right)^2\right),
\label{junction_conditiondim}  
\end{align}
and  boundary conditions at  infinity, which require that the  membrane stress returns to the isotropic value pre-compression (given by Laplace's equation) and that there is no rotation, i.e.
\begin{equation}
\Sigma(\infty)=\frac{pR}{2},\ \Psi(\infty)=0. 
\end{equation}

The hypothesis behind the kinematics are that strains are small but rotations may be moderate, i.e.~that the typical rotation angle is much smaller than unity, but much larger than the membrane strain; these are expressed in eqns (12.11b), (14.11) and (14.12) of \cite{audoly2010elasticity}, for example. (The resulting equations are not exact but are a truncation of the exact results with the elastic strain terms truncated following the linear term in displacement, while rotation terms retain terms quadratic in the rotation --- the smallest order term that includes the rotation.) The resulting equations also assume that the shell is ``shallow" and hence are valid while displacements are restricted to a zone of the shell that is well approximated by a parabola centered at the apex, i.e.
\begin{equation}
s\ll R,\quad s\ll R_o, \  z^\prime(s)\approx\frac{-s}{R}
\end{equation}
(There is no restriction on the indenter size, provided that the contact zone satisfies these requirements.)

The equations are made dimensionless by introducing the small parameter $\epsilon=(B/YR^2)^{1/2}$ in which $B=Eh^3/[12(1-\nu^2)]$ is the bending stiffness and $Y=Eh$ is the membrane stretching stiffness; hence $\epsilon=h/R/[12(1-\nu^2)]^{1/2}\ll1$ for a thin shell. The balance between the terms representing bending stresses and those representing curvature-induced in-plane stresses in \eqref{eq_Psidim} introduces the bending length $R\sqrt{\epsilon}\sim (hR)^{1/2}$ as a natural unit for the curvilinear coordinate $s$ \cite{Pogorelov1973}; we also use the shell radius $R$ as a unit of vertical displacement $w$, and $E\epsilon R h$ as the unit of force. The dimensionless quantities are then $\hat{s}={s}/{(R\sqrt{\epsilon})}$, $\hat{w}=w/R$, $\hat{\Psi}=\Psi/\sqrt{\epsilon}$,  $\hat{\Sigma}=\Sigma/(E h\epsilon)$, 
$\hat{p}=pR/(E\epsilon h)$, $\hat{\phi}=\phi/(E h R\epsilon)$, $\hat{F}=F/(E h R\epsilon)$. The compatibility  and  vertical force balance equations then take the form
\begin{equation}
3\hat{s}\hat{\Sigma}^\prime+\hat{s}^2\hat{\Sigma}^{\prime\prime}+\hat{s}\hat{\Psi}+\frac{\hat{\Psi}^2}{2}=0,\label{eq_Sigma}
 \end{equation}
 \begin{equation}
\hat{s}\hat{\Sigma}(\hat{s}+\hat{\Psi})-\frac{d(\hat{s}\hat{\Psi}^\prime)}{d\hat{s}}+\frac{\hat{\Psi}}{\hat{s}}=\hat{p}\frac{\hat{s}^2}{2}-\frac{\hat{\phi}}{2\pi}.\label{eq_Psi}
 \end{equation}
Continuity conditions at the periphery of the contact region are
\begin{equation}
\hat{\Psi}(\hat{s}_c)=-\hat{s}_c\left(1+\frac{R}{R_o}\right),\ \hat{\Psi}^\prime(\hat{s}_c)=-\left(1+\frac{R}{R_o}\right),\  \hat{\Sigma}^\prime(\hat{s}_c)=\frac{\hat{s}_c}{8}\left(1-\left(\frac{R}{R_o}\right)^2\right),
\label{junction_condition}
\end{equation}
and the boundary conditions at infinity become
\begin{equation}
\hat{\Sigma}(\infty)=\frac{\hat{p}}{2},\ \hat{\Psi}(\infty)=0. \label{infinity_condition}
\end{equation}
The system (\ref{eq_Sigma})-(\ref{infinity_condition}) is a fourth-order system with $\hat{F}$ determined as a function of $\hat{s}_c$ or vice-versa. The only other parameters in the problem are the dimensionless pressure, $\hat{p}$, and the ratio of shell to probe radius, $R/R_o$. 

The boundary value problem was solved using the ``bvp5c" solver of Matlab  which  provides $\hat{F}$ as part of the solution; the boundary conditions at  infinity were imposed by continuation by varying an end point \cite{cebeci1971shooting} (Figure \ref{Figure_2} a). As for a shell indented by a point tip, there is a linear relationship between force and displacement at small displacement (Figure \ref{Figure_2} b). The numerical method works well at higher $\tau$: the largest displacements that can be  obtained numerically range from a few tenths of the thickness at $\tau=0$ to ten times the thickness at $\tau=1000$; the numerical accuracy closely follow the accuracy of the initial guess provided by the analytical solution. In particular, for contact with a plane at high $\tau$, and with large displacements, we find that the membrane limit is retrieved, i.e.~the force scales as the area of the contact region multiplied by the pressure (data not shown). This reproduces the known membrane behavior of a pressurized membrane--shell, as found in the previous study by Feng \& Pangnan \cite{feng1973contact}.

\section{Small displacement behaviour}

\begin{figure}
\includegraphics[width=1\linewidth]{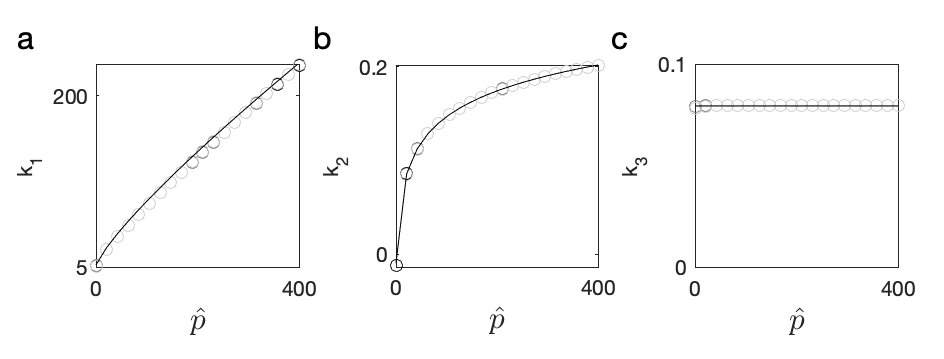}
\label{Figure_3}
\caption{Comparison of analytical and numerical solutions at small displacement. Panels (a), (b), and (c) show the constants $k_1$, $k_2$, and $k_3$ that are involved in the equations relating size of contact region, probe displacement, and force, respectively, plotted as functions of the dimensionless pressure, $\hat{p}$. Circles represent the constants estimated from fits to numerical solutions and while curves show the predictions of the analytical solution of the linearized equations. Black, dark gray, and gray circles stand for $R/R_o=0$ (a plane), $R/R_o=1$, and $R/R_o=10$, respectively. }
\end{figure}

For small displacement ($w \lessapprox  h$), we expand equations (\ref{eq_Sigma})-(\ref{eq_Psi}) around the solution at zero force, $\hat{\psi}=0$ and $\hat{\Sigma}=\hat{p}/2$. We linearize the equations in terms of $\hat{\Psi}$ and
$\hat{\mu}=\hat{s}(\hat{\Sigma}-\hat{p}/2)$ for $\hat{s}>\hat{s}_c$:
\begin{equation}
\hat{\Psi}=-\frac{d}{d\hat{s}}\left[\frac{1}{\hat{s}}\frac{d}{d \hat{s}}(\hat{s}\hat{\mu})\right]\label{eq_Sigma3}
\end{equation} 
\begin{equation}
\hat{\mu}\hat{s}+\hat{s}\hat{\Psi}\frac{\hat{p}}{2}-\frac{d(\hat{s}\hat{\Psi}^\prime)}{d\hat{s}}+\frac{\hat{\Psi}}{\hat{s}}=\frac{\hat{F}}{2\pi}.\label{eq_Psi3}
\end{equation}
Substituting (\ref{eq_Sigma3}) into (\ref{eq_Psi3}) yields a single fourth order equation for $\hat{\mu}$
\begin{equation}
\hat{\mu}-\frac{\hat{p}}{2}\frac{d}{d\hat{s}}\left[\frac{1}{\hat{s}}\frac{d}{d \hat{s}}(\hat{s}\hat{\mu})\right]+\frac{d}{d\hat{s}}\left[\frac{1}{\hat{s}}\frac{d}{d \hat{s}}\left(\hat{s}\left(\frac{d}{d\hat{s}}\left[\frac{1}{\hat{s}}\frac{d}{d \hat{s}}\left(\hat{s}\hat{\mu}\right)\right]\right)\right)\right]=\frac{\hat{F}}{2\pi\hat{s}}.\label{eq_complete}
\end{equation}
The analytical solution is a sum of the particular solution $\hat{F}/(2\pi\hat{s})$ and a solution of the homogeneous equation, i.e. a linear combination of the derivatives of $f_+(\hat{s})=K_0(\sqrt{\lambda_+}\hat{s})$ and
$f_-(\hat{s})=K_0(\sqrt{\lambda_-}\hat{s})$, $K_0$ being the zeroth order modified Bessel function of the second kind, and
\begin{equation}
    \lambda_{\pm}=\hat{p}/4\pm\sqrt{(\hat{p}/4)^2-1}.
\end{equation}
Accordingly, we may write (for $\hat{s}>\hat{s}_c$)
\begin{equation}
\hat{\mu}(\hat{s})=-\frac{a}{\lambda_+} f_+^\prime(\hat{s})-\frac{b}{\lambda_-} f_-^\prime(\hat{s})+\frac{\hat{F}}{2\pi\hat{s}}.\label{Sol_mu}
\end{equation}
Substitution in (\ref{eq_Sigma3}) yields
\begin{equation}
\hat{\Psi}(\hat{s})=af_+^{\prime}(\hat{s})+bf_-^{\prime}(\hat{s})\label{Solution_Psi}
\end{equation}
The boundary conditions at infinity are verified by the choice of $K_0(\cdot)$ and the constants $a$ and $b$ are obtained by imposing the boundary condition (\ref{junction_condition}) for $\hat{\Psi}$ and $\hat{\Psi}^\prime$. We find that
\begin{equation}
a=\left(1+\frac{R}{R_o}\right)[f_-^{\prime}(\hat{s}_c)-\hat{s}_cf_-^{\prime\prime}(\hat{s}_c)][f_+^{\prime}(\hat{s}_c)f_-^{\prime\prime}(\hat{s}_c)-f_-^{\prime}(\hat{s}_c)f_+^{\prime\prime}(\hat{s}_c)]^{-1}
\end{equation}
\begin{equation}
b=\left(1+\frac{R}{R_o}\right)[f_+^{\prime}(\hat{s}_c)-\hat{s}_c f_+^{\prime\prime}(\hat{s}_c)][f_+^{\prime}(\hat{s}_c)f_-^{\prime\prime}(\hat{s}_c)-f_-^{\prime}(\hat{s}_c)f_+^{\prime\prime}(\hat{s}_c)]^{-1}
\end{equation}
The boundary condition (\ref{junction_condition}) for $\hat{\Sigma}^\prime$ yields a relation between $\hat{F}$ and $\hat{s_c}$:
\begin{align}
\hat{F}&=\left(1+\frac{R}{R_o}\right)\pi\frac{(f_+^{\prime}(\hat{s}_c)-\hat{s}_cf_+^{\prime\prime}(\hat{s}_c))(f_-^{\prime}(\hat{s}_c)-\hat{s}_cf_-^{\prime\prime}(\hat{s}_c))\sqrt{(\hat{p}/4)^2 - 1}}{f_+^{\prime}(\hat{s}_c)f_-^{\prime\prime}(\hat{s}_c)-f_-^{\prime}(\hat{s}_c)f_+^{\prime\prime}(\hat{s}_c)}\nonumber\\&+\pi\left(1-\frac{R^2}{R_o^2}\right)\frac{\hat{s}_c^4}{8}.\label{Solution_F}
\end{align}

Finally we obtain the main physical quantities. From Eq. \ref{Sol_mu}, the meridional membrane stress is
\begin{equation}
\hat{\Sigma}_{ss}=\frac{\hat{p}}{2}-\frac{\hat{F}}{2\pi \hat{s}^2}-\frac{a}{\hat{s}\lambda_+} f_+^\prime(\hat{s})-\frac{b}{\hat{s}\lambda_-} f_-^\prime(\hat{s})\label{Solution_Sigma}.
\end{equation}
Following formula (14.20) of \cite{audoly2010elasticity}, the dimensionless membrane circumferential stress is
\begin{equation}
\hat{\Sigma}_{\theta\theta}=\frac{\hat{p}}{2}+\frac{\hat{F}}{2\pi \hat{s}^2}-\frac{a}{\lambda_+} f_+^{\prime\prime}(\hat{s})-\frac{b}{\lambda_-} f_-^{\prime\prime}(\hat{s})
\end{equation}
The geometrical displacement is obtained by integrating (\ref{Solution_Psi}):
\begin{equation}
\hat{w}(\hat{s})=-af_+(\hat{s})-bf_-(\hat{s}).
\end{equation}

To obtain simple analytical formulae, we further consider the limit of a small contact region (relative to the bending length) i.e.~$\hat{s}_c\ll1$ --- this is an approximation that holds for most of the range of simulations shown above. We also return to dimensional quantities, recalling that the dimensionless parameters
\begin{eqnarray}
    \epsilon&=&\frac{h}{[12(1-\nu^2)]^{1/2}R}\label{epsilon} \\ 
    \hat{p}&=&\frac{pR}{\epsilon E h} \label{phat}
\end{eqnarray} with $\epsilon$ in particular used in the definition of the horizontal length scale.

The relation between the vertical displacement at the origin $w_0$ and $s_c$, the radius of the contact region, is given  by
\begin{equation}
    w_0=R\epsilon\frac{1+R/R_o}{k_1\bigl[k_2-k_3\log(s_c/(R\sqrt{\epsilon}))\bigr]},
    \label{Formula_displacement}
\end{equation} where we introduce three constants
\begin{equation}
    k_1=\frac{4\pi\sqrt{(\hat{p}/4)^2 - 1}}{\arctanh(\sqrt{1-(\hat{p}/4)^{-2}})},\label{eq:k1}
\end{equation}
\begin{equation}
    k_2=\frac{\hat{p}}{8k_1}+ \frac{\gamma-\log(2)}{4\pi},\label{eq:k2}
\end{equation}
\begin{equation}
   k_3=1/4\pi, \label{eq:k3}
\end{equation}
 and

where $\gamma$ is Euler's $\gamma$ constant. The first and third constants have the following limiting behaviours as a function of dimensionless pressure: $k_1\sim 8$ and $k_2\sim \frac{\gamma-\log(2)}{4\pi}$ for small $\hat{p}$, while $k_1\sim \pi \hat{p} /\log\hat{p}$ and $k_2\sim 1/(8\pi)\log\hat{p}$ for large $\hat{p}$. We manually got some point on the curve ($w$ vs $F/(2\pi)$) of (\cite{audoly2010elasticity}) (Figure 14.15): the slope at the beginning is $1.39$ which times $2\pi$ gives $8.75$ close from  $k_1=8$. Note, in particular, that the expression \eqref{eq:k1} is well-defined and real for $\hat{p}<4$.

The relation between the force and $s_c$ is given by
\begin{equation}
    F=\epsilon^2 E R h\frac{1+R/R_o}{k_2-k_3\log(s_c/(R\sqrt{\epsilon}))} \label{Formula_force}
\end{equation} 
Taking the ratio of (\ref{Formula_force})  and   (\ref{Formula_displacement}) leads to a considerable simplification; at lowest order in $\hat{s}_c$, the force--displacement relationship is linear:
\begin{equation}
F=\epsilon E h k_1 w_0.\label{functional_form_F_w}
\end{equation}
Surprisingly, this equation does not involve the probe size, $R_o$. Indeed, the lowest order term (in $\hat{s}_c$) of $F/w$ that depends on $R/R_o$ is
\begin{equation}
-\epsilon E h\pi\left(1-\frac{R}{R_o}\right)\frac{k_1k_3\log(s_c/(R/\sqrt{\epsilon}))(s_c/(R/\sqrt{\epsilon}))^4}{8}
\end{equation}
which remains negligible for small $\hat{s}_c$.

To compare these results with our numerical results, we fitted numerical solutions (Figure \ref{Figure_3} a-c) to the functional forms (\ref{functional_form_F_w},\ref{Formula_force},\ref{Formula_displacement}), with $k_1$, $k_2$, and $k_3$ taken as free parameters, and we compared these values to the theoretical values Eqs.~(\ref{eq:k1}-\ref{eq:k3}). Consistent with our approximation, the match is good for small displacements when the pressure is high enough (Figure \ref{Figure_3} a-c). The $w$ vs $F$ curve is independent of $R_o/R$ at lowest order, explaining that the dimensionless stiffness $k_1$ is identical to that obtained for a point indenter \cite{vella2012indentation}.

\section{Conclusion}
We find that, at small compression, the pressurized spherical shell behaves as a linear spring with stiffness given by 
$$K=\epsilon E h \frac{4\pi\sqrt{(\hat{p}/4)^2 - 1}}{\arctanh(\sqrt{1-(\hat{p}/4)^{-2}})}$$ 
(with $\epsilon$ and $\hat{p}$ defined by Eqs. \ref{epsilon}-\ref{phat})
independently of the probe geometry. This result is potentially useful for the wide range of  experimental protocols mentioned in the introduction. At low pressure, stiffness $K$ is mostly sensitive to modulus and at high pressure stiffness is mostly sensitive to pressure. Therefore, experiments in both normal medium and high-osmolarity media make it possible to deduce modulus and pressure (assuming Poisson's ratio has a standard value). 
If the monitoring of the contact area is possible, then the comparison of observed contact radius with theoretical radius at a given force $F$
$$s_c=\sqrt{\epsilon} R \exp\left[4\pi(k_2-\epsilon^2 E R h(1+R/R_o)/F)\right]$$
(with $\epsilon$ and $\hat{p}$ defined by Eqs. \ref{epsilon}-\ref{phat} and $k_3$ defined by Eq. \ref{eq:k3}) provides a simple test of whether it is appropriate to model the sample as a pressurized shell.

There are an increasing number of studies on mechanoperception and mechanotransduction in living systems \--- see reviews about plants~\cite{Codjoe.2021}, animals~\cite{Vining.2017,Mao.2016}, fungi~\cite{Davi.2015}, or bacteria~\cite{Sun.2011}. These studies call for theoretical descriptions of how external probes affect force distributions in cells and tissues. Indeed, the simplest experiment on invasive growth by plant roots or fungal hyphae involves putting the growing body in front of a plane and measuring the force upon impact (\cite{bizet20163d},\cite{money2001biomechanics}). The deformation at the tip triggers a growth response: a quantitative study of this response first requires a precise description of tip deformation at contact, for which our study provides a first basis.

\section{Funding statement}
Etienne Couturier is funded by ANR AnAdSPi (ANR-20-CE30-0005). Arezki Boudaoud is funded by  ANR-17-CE20-0023-02 WALLMIME, ANR-20-CE13-0022-03 HydroField, ANR-20-CE13-0003-02 CellWallSense.

\section{Author contribution statement}
EC, DV and AB designed the research. EC perfomed the research. EC, DV and AB wrote the article.

\printbibliography

\end{document}